\documentclass{aa}  

\usepackage{graphicx}
\usepackage{txfonts}
\newcommand{\equ}[1]{eq.~(\ref{eq:#1})}
\newcommand{\equs}[1]{eqs.~(\ref{eq:#1})}
\newcommand{\equm}[1]{(\ref{eq:#1})}
\newcommand{\Equ}[1]{Eq.~(\ref{eq:#1})}

\newcommand{\se}[1]{\S\ref{sec:#1}}
\newcommand{\fig}[1]{Fig.~\ref{fig:#1}}

\newcommand{\Fig}[1]{Figure~\ref{fig:#1}}

\newcommand{\be}{\begin{equation}}
\newcommand{\ee}{\end{equation}}
\newcommand{\bea}{\begin{eqnarray}}
\newcommand{\eea}{\end{eqnarray}}

\newcommand{\msun}{{\rm M}_\odot}
\newcommand{\Msun}{M_\odot}

\newcommand{\ifm}[1]{\relax\ifmmode#1\else$\mathsurround=0pt #1$\fi}
\newcommand{\kms}{\ifmmode\,{\rm km}\,{\rm s}^{-1}\else km$\,$s$^{-1}$\fi}

\newcommand{\kpc}{\,{\rm kpc}}

\newcommand{\Myr}{\,{\rm Myr}}

\newcommand{\ltsima}{$\; \buildrel < \over \sim \;$}
\newcommand{\lsim}{\lower.5ex\hbox{\ltsima}}
\newcommand{\gtsima}{$\; \buildrel > \over \sim \;$}
\newcommand{\gsim}{\lower.5ex\hbox{\gtsima}}

\def\Mv{M_{\rm vir}}

\def\Ms{M_*}

\usepackage{color}

\begin{document}

   \title{Redshift-dependent galaxy formation efficiency at $z=5-13$ in the FirstLight Simulations}

   \author{D. Ceverino
          \inst{1,2}, 
          Y. Nakazato \inst{3}, N. Yoshida  \inst{3,4,5}, R. S. Klessen \inst{6,7}, 
          \and
          S. C. O. Glover \inst{6}
          }

   \institute{Departamento de Fisica Teorica, Modulo 8, Facultad de Ciencias, Universidad Autonoma de Madrid, 28049 Madrid, Spain\\
              \email{daniel.ceverino@uam.es} 
         \and
             CIAFF, Facultad de Ciencias, Universidad Autonoma de Madrid, 28049 Madrid, Spain
         \and
         Department of Physics, The University of Tokyo, 7-3-1 Hongo, Bunkyo, Tokyo 113-0033, Japan 
         \and 
         Kavli Institute for the Physics and Mathematics of the Universe (WPI), UT Institute for Advanced Study, The University of Tokyo, Kashiwa, Chiba 277-8583, Japan
         \and
         Research Center for the Early Universe, School of Science, The University of Tokyo, 7-3-1 Hongo, Bunkyo, Tokyo 113-0033, Japan
         \and
         Universit\"{a}t Heidelberg, Zentrum f\"{u}r Astronomie, Institut f\"{u}r Theoretische Astrophysik, Albert-Ueberle-Str. 2, 69120
Heidelberg, Germany
	\and
	Universit\"{a}t Heidelberg, Interdisziplin\"{a}res Zentrum f\"{u}r Wissenschaftliches Rechnen, INF 205, 69120, Heidelberg, Germany\\
             }

   \date{Received Month day, year; accepted Month day, year}

\titlerunning{Variable efficiency}
\authorrunning{Ceverino et al.}

 
  \abstract
   {Some models of the formation of first galaxies predict low masses and faint objects at extremely high redshifts, $z\simeq9-15$. 
   However, the first observations of this epoch indicate a higher-than-expected number of bright (sometimes massive) galaxies.
   }
   {Numerical simulations can help to elucidate the mild evolution of the bright end of the UV luminosity function 
   and they can provide the link between the evolution of bright galaxies and variations of the galaxy formation efficiency across different redshifts.}
   {We use the FirstLight database of 377 zoom-in cosmological simulations of a volume- and mass-complete sample of galaxies.
   Mock luminosities are estimated by a dust model constrained by observations of the  $\beta$-M$_{\rm UV}$ relation at $z=6-9$.}
   {FirstLight contains a high number of bright galaxies, M$_{\rm UV} \le -20$, consistent with current data at $z=6-13$.
   The  evolution of the UV cosmic density is driven by the evolution of the galaxy efficiency and the relation between  M$_{\rm UV}$ and halo mass.
The efficiency of  galaxy formation increases significantly with mass and redshift. At a fixed mass, galactic halos at extremely high redshifts convert gas into stars at a higher rate than at lower redshifts. The high gas densities in these galaxies enable high efficiencies.
Our simulations predict higher number densities of  massive  galaxies, $\Ms \simeq 10^{9} \ \Msun$, than other models with constant efficiency. }
   {Cosmological simulations of galaxy formation with detailed models of  star formation and feedback can reproduce the different regimes of galaxy formation across cosmic history. }


   \maketitle
   
      \nolinenumbers  
      
%

\section{Introduction}

The formation of the first galaxies marks the end of the cosmic dark ages and the beginning of cosmic dawn.
The Universe during this first-light epoch was dense, mostly neutral and pristine.
The conditions for galaxy and star formation were very different than at later times \citep{Abel02, Bromm02, Yoshida08, KlessenGlover23}.
Thanks to the brand new {\it James Webb} Space Telescope (JWST), we are now getting a first glimpse of this epoch.

The most surprising result from JWST is the higher-than-expected number of bright (sometimes massive) galaxies observed at extremely high redshifts, $z=9-15$ \citep{Naidu22, Castellano22, Adams23, Atek23, Donnan23, Harikane23, PerezGonzalez23, Leung23, Hainline24, Casey24, Finkelstein23,Yan23}. 
These galaxies are bright in the rest-frame UV, M$_{\rm UV}<-20$, and they are relatively massive, $\Ms=10^8-10^9 \ \Msun$, for that early epoch. 
As a result, both cosmic UV and mass density decrease gently with redshift and this pushes the formation of first galaxies to even higher redshifts.
These observations, although uncertain, challenge our understanding of the growth of galaxies at these early times.

Several solutions have been proposed.
For example, UV variability due to bursts of star formation (SF) could impact the bright-end of the UV luminosity function (UVLF) at these high redshifts  \citep{Shen23, Sun23, Gelli24}. However, it is not clear whether the necessary degree of burstiness  is consistent with the SF histories of these galaxies. 
\cite{Mason23} similarly argue that current observations are biased toward the most extreme star-forming galaxies for a given halo mass. 
This highlights the need for deeper observations of galaxies at $z>9$.
If dust attenuation is negligible at  high-z, UV bright galaxies can be more abundant than expected \citep{Ferrara23, Tsuna23}. 
A related solution involves the evolution of the initial mass function (IMF). A top-heavy IMF due to Pop III stars can boost the UV luminosities by a factor of a few \citep{Yung23, Yung24}.
However, metallicity increases quickly after the first supernovae (SN) explosions and these SN shells produce dust in short time-scales \citep{Lesniewska19}. Therefore, the relatively massive galaxies observed by JWST should also host significant amounts of metals and dust.
Some of these stellar mass estimates show a 3-$\sigma$ tension with the $\Lambda$CDM scenario, but this does not necessarily falsify the current paradigm \citep{Lovell23}.
Instead, better mass estimates from photometry fitting are needed to constrain complex SF histories and total stellar masses. 

The intrinsic UV luminosities of these galaxies are related to their star formation rates (SFRs) and galaxy growth.
The fuel for SF is provided mainly by gas accretion into galaxies \citep{Dekel09}. 
At a first approximation, this mass accretion is regulated by the halo growth \citep{Bouche10, Lilly13, Dekel13}. We can therefore define a galaxy formation efficiency as the ratio between stellar and halo growth. 
This definition excludes  the growth of the gaseous component \citep{Dekel13}, as these UV luminosities mostly trace stellar light.
In principle this efficiency may vary with mass and redshift, depending on the particular conditions of the galaxy/halo assembly. A higher efficiency  at extremely high redshifts could explain the excess of UV-bright galaxies at $z\geq9$.
 \cite{Dekel23} propose a scenario in which stellar feedback from massive stars is not able to regulate the SF process in massive galaxies at high-z.
  In the Feedback Free Bursts (FFB) model, the SF efficiency is higher than in starbursts at lower redshifts  because of the  higher gas densities in star-forming regions.
 \cite{LiDekel23} provides predictions that can be compared with current observations.

Cosmological simulations that predict the efficiency of feedback need to resolve the small scales where mass, energy and momentum are injected into the interstellar medium \citep{Ceverino09,Fichtner24}. Only zoom-in simulations resolve the expansion of over-pressured bubbles that merge into galactic-scale outflows. However, most of these simulations only sample a few regions \citep{Oshea15, Ma18, Katz19, Pallottini19} and therefore it is difficult to make predictions of the overall galaxy population. The FirstLight database \citep{PaperI, PaperII} is a mass-complete sample of zoom-in simulations that allows population-averaged statistics with 10-20 pc resolution in a $(60 {\rm Mpc})^3$ comoving volume.
For example, \cite{PaperII} characterize the typical SF bursts at these high redshifts, using more than 1000 individual starbursts.

 A subsample of the FirstLight simulations shows that the stellar-to-halo mass relation evolves by a factor 3 between $z=6$ and $z=10$ for a fixed halo mass around $\Mv\sim10^{10} \msun$ \citep{PaperI}.
This evolution is related to the galaxy formation efficiency. 
In this paper we extend that relation to higher masses and redshifts.
We investigate whether the FirstLight simulations with  a diverse feedback model (thermal+radiative+kinetic) is able to reproduce the JWST observations of bright  galaxies at $z\ge9$ as well as  previous HST observations at lower redshifts, $z\simeq6$. This paper also links the origin of the evolution of the UVLF with a redshift-dependent galaxy formation efficiency. 
The outline of this paper is as follows.
Section \se{runs} summarizes the FirstLight simulations. 
Section \se{dust} computes the UV luminosity, taking into account a redshift-dependent dust attenuation model.
Section \se{results} provides the main findings, Section \se{discussion} discusses the results and Section \se{conclusions} provides a summary and a general conclusion.  

   \begin{figure*}
   \includegraphics[width= \columnwidth]{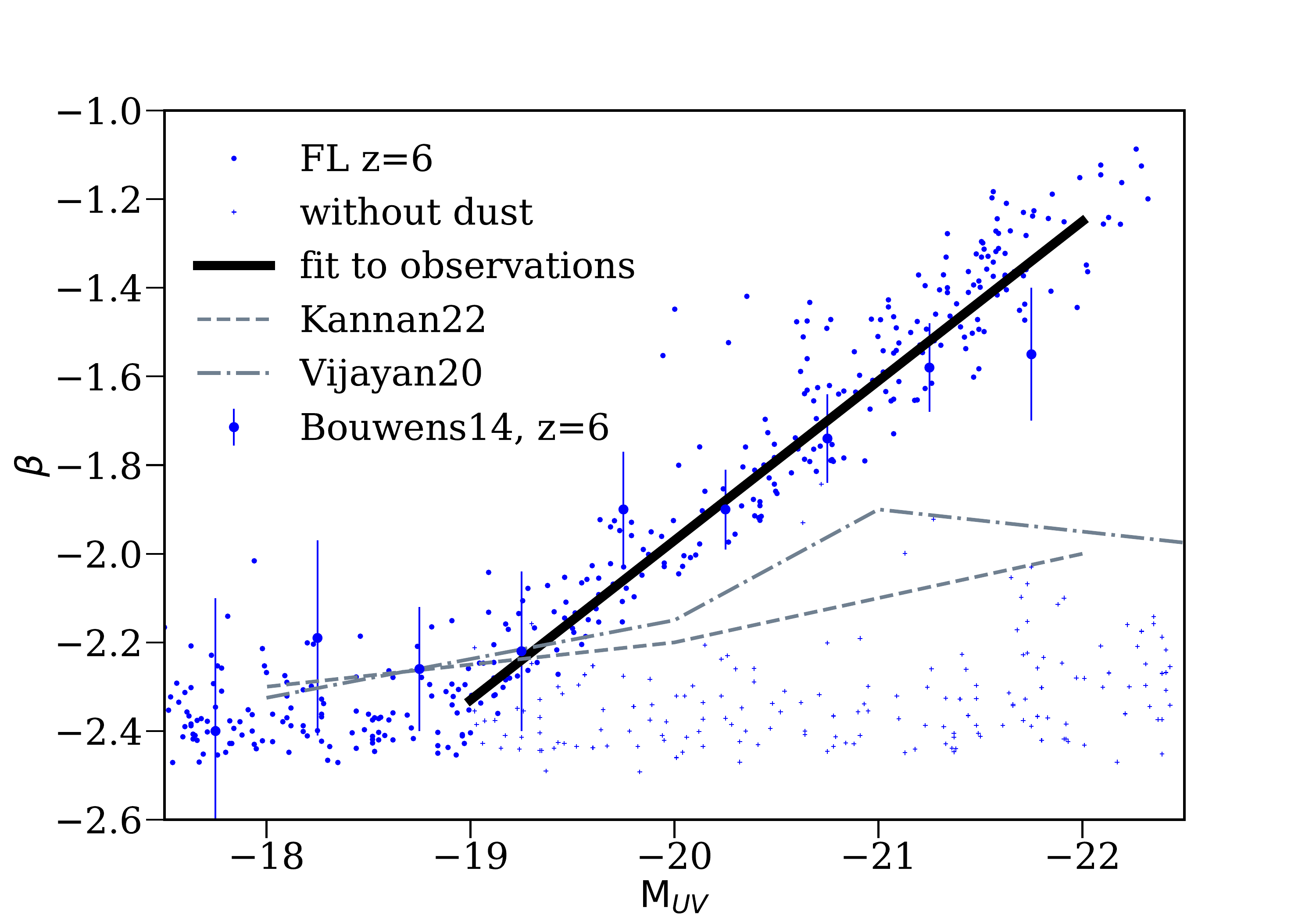}
   \includegraphics[width= \columnwidth]{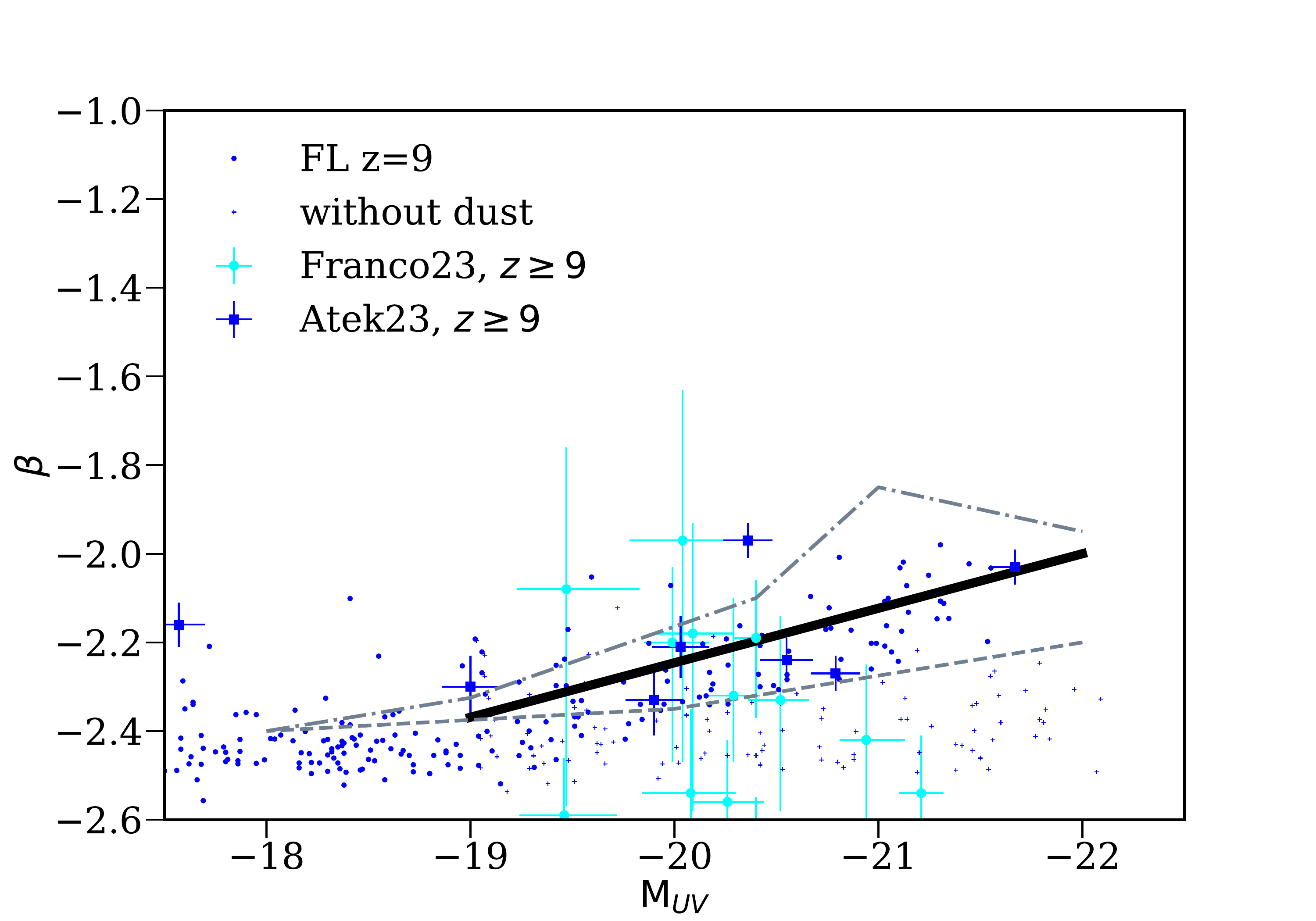}
   \caption{Constrained relations between the UV continuum slope, $\beta$, and the absolute UV magnitude at $z=6$ (left) and $z=9$ (right).
   Small crosses represent intrinsic values.
   Points represent dust-attenuated values. The solid black lines fit observations by \citet{Bouwens14} at $z=6$ and by \citet{Atek23} at $z\ge9$. 
   Grey dashed and dash-dotted  lines show models by \citet{Vijayan21} and  \citet{Kannan22} that fail to reproduce these observations. }
	  \label{fig:dust}
    \end{figure*}
%
  
%
%

\section{The FirstLight simulations}
\label{sec:runs}

The FirstLight simulations are multi-object, zoom-in cosmological simulations of a mass-complete sample of galaxies. 
The initial conditions are first described in  \cite{PaperI}.
The suite is composed by three cosmological boxes.
The 10-Mpc/h and 20-Mpc/h samples contain all halos within these volumes with a maximum circular velocity, V$_{\rm max}>50 \kms$ and  V$_{\rm max}>100 \kms$ respectively at $z=5$.
Similarly, the 40-Mpc/h box contains galaxies with  V$_{\rm max}>180 \kms$. 
Overall, the  FirstLight database includes 377 galaxies.

The simulations are performed with the  $N$-body+Hydro \textsc{ART} code
\citep{Kravtsov97,Kravtsov03, Ceverino09, Ceverino14, PaperI}.
Gravity and hydrodynamics are solved by an Eulerian, adaptive mesh refinement (AMR) approach.
The code includes  astrophysical processes relevant for galaxy formation, such as gas cooling by hydrogen, helium and metals.
Photoionization heating uses a redshift-dependent cosmological UV background with partial self-shielding. 

Star formation and feedback (thermal+kinetic+radiative) models are described in \cite{PaperI}.
In short, star formation is assumed to occur at densities above a threshold of 1 cm$^{-3}$ and at temperatures below $10^4$ K. The code implements a stochastic star formation model that scales with the gas free-fall time \citep{Schmidt, Kennicutt98}.
In addition to thermal energy feedback, the simulations use radiative feedback, as a local approximation of radiation pressure. This model adds non-thermal pressure to the total gas pressure in regions where ionizing photons from massive stars are produced and trapped. The model of radiative feedback is named RadPre\_IR in \cite{Ceverino14} and it uses a moderate trapping of infrared photons. 
The kinetic feedback model also includes the injection of momentum coming from the (unresolved) expansion of gaseous shells from supernovae and stellar winds \citep{OstrikerShetty11}.  More details can be found in \cite{PaperI}, \cite{Ceverino14}, \cite{Ceverino10}, and \cite{Ceverino09}. 

For the 10 and 20 Mpc boxes, the DM particle mass resolution is $m_{\mathrm{DM}} = 10^4 \ \msun$ and  the minimum mass of star particles is $10^2 \ \msun$.  The maximum spatial resolution is between 8.7 and 17 proper pc (a comoving resolution of 109 pc after $z<11$). For the 40-Mpc box, the spatial and mass resolution  are two and eight times lower respectively.
 
 Previous papers using this database predict rest-frame spectral energy distributions (SEDs), and emission from visible  \citep{PaperIII, PaperIV} and far-infrared \citep{Nakazato23} lines.  
 In addition, FirstLight predicts a weak evolution of the mass-metallicity relation at $z\ge5$ \citep{Langan20}, consistent with current findings \citep{Curti23, Venturi24}.


\section{Dust attenuation and evolution of the $\beta$-M$_{\rm UV}$ relation}
\label{sec:dust}

   \begin{figure*}
   \centering
   \includegraphics[width=2 \columnwidth]{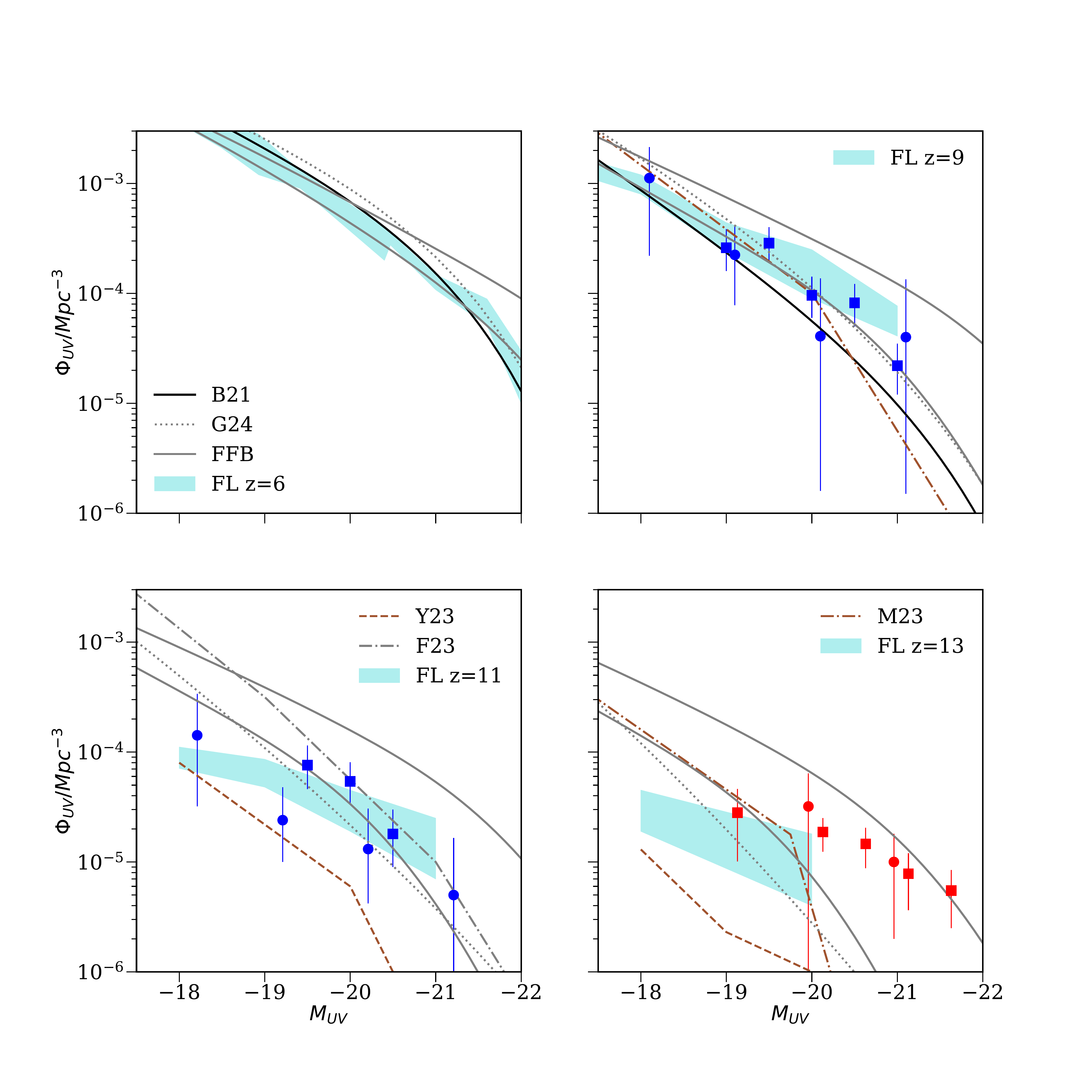}
   \caption{Evolution of the UV luminosity function from $z=6$ to $z=13$. The cyan regions represent the FirstLight simulations with Poisson uncertainties. JWST observations are marked by blue circles \citep{Harikane23}, blue squares \citep{Finkelstein23}, red circles \citep{Bouwens22}, and red squares \citep{Yan23}. Pre-JWST observations are shown by black lines \citep{Bouwens21}. The fiducial models by \citet{Ferrara23} (grey  dash-dotted line), \citet{Yung23} (brown dashed lines),  \citet{Mauerhofer23} (brown dash-dotted lines), and \citet{Gelli24} (gray dotted lines) fail to reproduce JWST observations at $z\ge9$. FirstLight results lie in between FFB models with $\epsilon_{\rm max}=0.2$ and 1 (gray solid lines). }
	  \label{fig:UVLF}
    \end{figure*}

The rest-frame UV luminosities can be affected by dust attenuation. This reduces the observed flux but also modifies the UV slope, $\beta$. 
In order to compare to observations, we include a dust attenuation model constrained by current observations that show an evolution of $\beta$ with redshift \citep{Bouwens14, Atek23, Roberts-Borsani24}. They generally suggest that dust attenuation is lower at higher redshifts.

The intrinsic SEDs of FirstLight galaxies are described in \citet{PaperIII}.
The stellar spectrum from 1 to 100 000 Å coming from each star particle uses the templates of single stellar populations (SSP) from the Binary Population and Spectral Synthesis (BPASS) model \citep{Eldridge17} including nebular  emission \citep{XiaoStanway18}.
BPASS v2.1 assumes a Kroupa-like initial mass function with power slopes $ \alpha=-1.3$ for star masses $m = 0.1-0.5  \ \msun$ and 
$\alpha_2=2.35$ for star masses $m = 0.5-100 \ \msun$. 
From these SEDs, we compute the dust-free values of M$_{\rm UV,0}$ and $\beta_{\rm 0}$. 
The UV absolute magnitude is defined at 1500 \AA \ with a bandwidth of 300 \AA.
The UV slope is computed between 1700 and 2200 \AA \ \citep{Bouwens14}.

In \citet{PaperIII}, we compare these intrinsic values with the observed relation at $z\sim6$ by \citet{Bouwens14}. 
At magnitudes brighter than M$_{\rm UV,0}=-19$, there is a mismatch between  observed and intrinsic values due to dust attenuation. We fit these observed values with a mean relation $\langle \beta \rangle=f(M_{\rm UV})$ following the observations at  $z\sim6$ by \citet{Bouwens14} and by \citet{Atek23} at $z\geq9$ (\Fig{dust}).

For galaxies with M$_{\rm UV,0}<-19$, we compute the ($\beta$, M$_{\rm UV}$) pair needed to follow these observed relations, starting with the intrinsic values ($\beta_{0}$, M$_{\rm UV,0}$). First, we compute the average intrinsic slope, $\langle \beta_{0} \rangle=-2.37$, for all galaxies brighter than  M$_{\rm UV,0}=-19$. For each galaxy, we compute the deviation from this average, $\Delta \beta_{0}= \beta_{0} \ - \langle \beta_{0}\rangle$. This scatter is kept  invariant because it is due to the intrinsic differences in the age of the stellar populations.
Then, a first guess of the attenuated value is
\begin{equation} 
\beta= f(x)+ \ \Delta \beta_{0},
\label{eq:f}
\end{equation}
where $x={\rm M}_{\rm UV,0}$, the intrinsic magnitude.
The dust attenuation, A$_{\rm UV}$, is computed using the definition of the UV slope and the relation between intrinsic and attenuated values:
\begin{equation} 
\beta=\beta_0 - C \ {\rm A}_{\rm UV},
\end{equation}
where $C=-0.55$ assuming a Calzetti attenuation law \citep{Calzetti}. Therefore, the attenuated magnitude is M$_{\rm UV}$=M$_{\rm UV,0} +$A$_{\rm UV}$. 
Using this value, we recompute $\beta$ using \Equ{f} with $x={\rm M}_{\rm UV}$. The new $\beta$ is now consistent with the dust attenuated magnitude. 


\Fig{dust} shows the $\beta$-M$_{\rm UV}$ relations from FirstLight galaxies, constrained by observations.
Other attempts to compute dust attenuation using radiative transfer calculations \citep{Vijayan21,Kannan22} are not able to reproduce relatively red slopes, $\beta>-1.8$, which are very common in bright sources, M$_{\rm UV}<-20$ at $z\simeq6$ \citep{Bouwens14}.
Our approach to dust attenuation reproduces these red slopes by construction. Future attempts using radiative transfer calculations will compute dust attenuation self-consistently from galaxy properties. 
This requires high-resolution simulations that resolve the dust-star mixture accurately \citep{Mushtaq, Esmerian23}, as well as the holes and clumps in the dust distribution.

At higher redshifts, $z\ge9$, JWST observations indicate lower attenuations at a fixed UV magnitude \citep{Atek23, Franco23, Heintz24}.
This could be due to a lower dust or metal content in these first galaxies. Our estimation of dust attenuation is constrained by these observations but larger samples of galaxies with accurately values of $\beta$ at $z\geq9$ are needed to confirm these trends.

\section{Results}
\label{sec:results}

\subsection{Evolution of the UV luminosity function}

Using the dust-attenuated rest-frame UV absolute magnitudes constrained by the observed evolution of the  $\beta$-M$_{\rm UV}$ relation between $z=6$ and $z\ge9$, we can generate the UV luminosity functions at different redshifts and compare with pre-JWST and JWST observations (\Fig{UVLF}).
At $z=6$, FirstLight galaxies are consistent with the compilation of pre-JWST observations by \citet{Bouwens21}, although the faint end falls slightly below observations by about 0.1 dex.

At $z=9$, The FirstLight simulations are consistent with new JWST observations \citep{Harikane23,Finkelstein23}. The number densities for bright sources,  M$_{\rm UV}<-20$, are higher than in pre-JWST data.
This trend continues at higher redshifts, $z=11$ and 13, where FirstLight number densities are remarkably similar to the observed JWST estimates \citep{Bouwens22, Yan23}, although the uncertainties are high for these relatively small samples. 
In particular, FirstLight lacks statistics for very bright galaxies,  M$_{\rm UV}<-20$ at the highest redshift bin, $z=13$. Most probably this is due to the relatively small volume of these simulations. On the other hand, the two galaxies with spectroscopic redshifts  at $z=13-14$ are not exceptionally  bright, M$_{\rm UV}>-21$ \citep{Carniani24}. More spectroscopic redshifts of high-z candidates are needed for a better constrain of the high-luminosity end of the  UVLF at these extremely high redshifts.

FirstLight predictions are higher than recents models that assume a constant galaxy efficiency \citep[e.g.][]{Yung23, Mauerhofer23, Gelli24}, especially for bright galaxies at $z=13$.
Other models, such as \citet{Ferrara23}, overproduce fainter galaxies in comparison with FirstLight and JWST results.
The closest model is  FFB  with $\epsilon_{\rm max}=0.2$ \citep{Dekel23,LiDekel23}.

Bright galaxies form earlier and faster than fainter galaxies at $z\ge9$ and therefore, their number density increases slowly at later times.
For example, the number density of galaxies with M$_{\rm UV}=-19$ increases by two orders of magnitudes from $z=13$ to $z=6$. During the same period of time, the number density of galaxies with M$_{\rm UV}=-20$ increases only by a factor of 60. 
For a better understanding of this galaxy growth, we need to link these UV magnitudes with the virial mass of the halos hosting these galaxies. 
According to gas-regulator models \citep{Bouche10, Lilly13, Dekel13, Dekel14},
the halo mass at a given redshift sets the amount of gas accreted into the galaxy and regulates the subsequent star formation at the central galaxy.  Then, we can look at the galaxy growth at different halo masses and redshifts.


\subsection{Scaling relations: $\Mv$ vs M$_{\rm UV}$ }

   \begin{figure}
   \centering
   \includegraphics[width= \columnwidth]{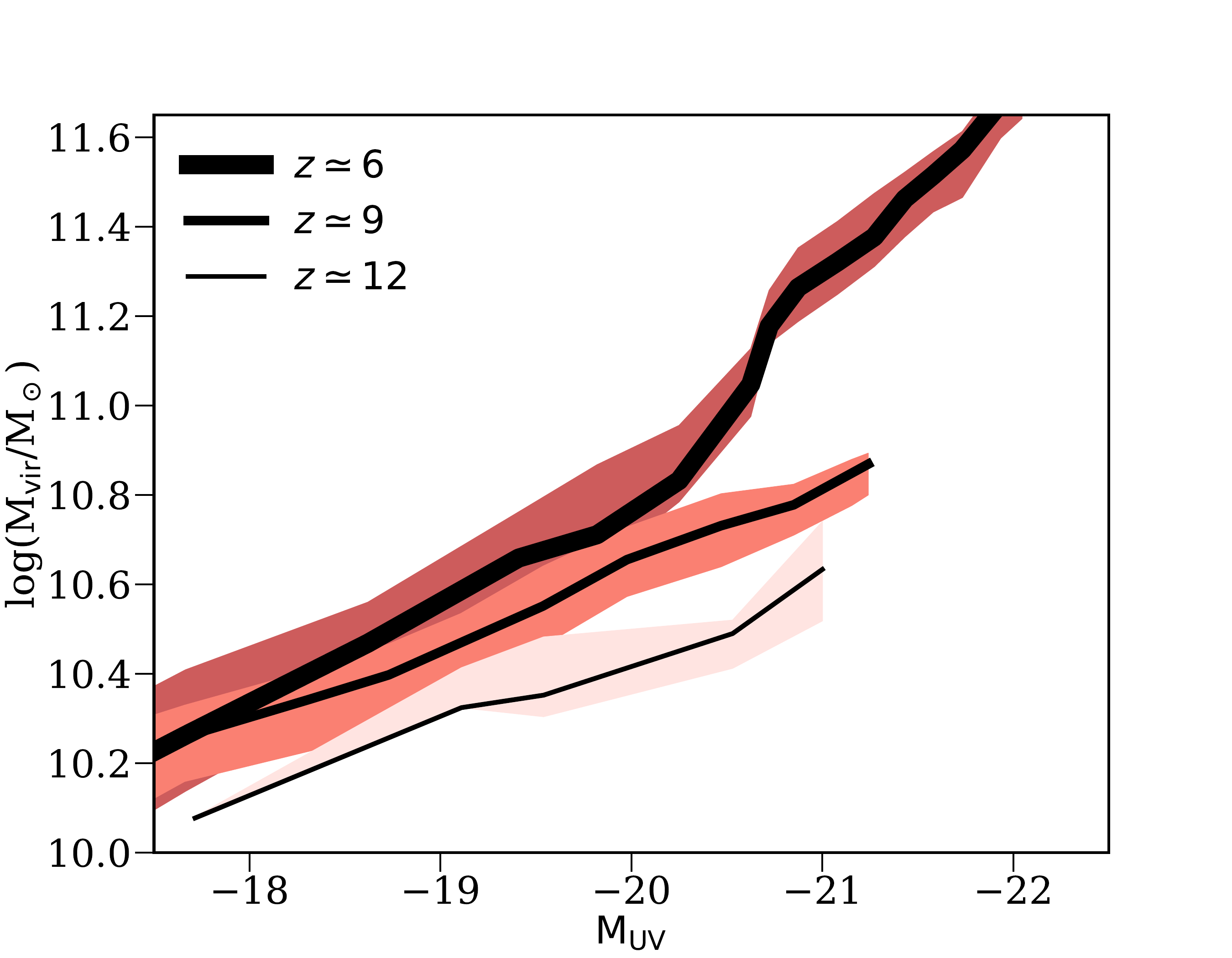}
   \caption{Virial mass versus M$_{\rm UV}$ at $z\simeq6$, 9 and 12 in FirstLight galaxies.
      The colored regions cover the 25 and 75\% quantiles. 
As redshift increases, the UV luminosity increases at a fixed virial mass.
}
              \label{fig:scalingR}%
    \end{figure}

The relation between the UV magnitude and the stellar (or halo) mass evolves with redshift \citep{PaperIII}. This is due to the evolution of the specific star-formation rate (sSFR), partially driven by the specific halo mass growth, $\dot{M}_{\rm vir}/\Mv \propto (1+z)^{5/2}$ \citep{Dekel13,Dekel14,PaperII}.
\Fig{scalingR} shows the evolution of the relation between the UV magnitude and the virial mass, $\Mv$ \citep{BryanNorman}.
A bright galaxy with  M$_{\rm UV}=-20$ at $z\simeq6$,
corresponding to an Universe age of $t_{\rm U}\simeq1 \ {\rm Gyr}$, typically lives in a halo of mass $ {\rm log}(\Mv/\msun)=10.8 \pm 0.1$.
A second galaxy with the same rest-frame UV magnitude but at $z=12$ ($t_{\rm U}\simeq 0.4 \ {\rm Gyr}$) has a halo of mass $ {\rm log}(\Mv/\msun)\simeq10.4$, a factor three lower than in the first case.
Galaxies with masses similar to the second halo are therefore more abundant than galaxies with the first halo mass at $z=12$.
As a result, the bright sources have higher number densities at $z=12$ than in the case of no evolution in the 
 M$_{\rm UV}$-$\Mv$ relation.

For brighter objects, M$_{\rm UV}<-20$, there is a change in the slope of the relation due to dust attenuation at $z=6$. 
This increases the difference described above between galaxies with M$_{\rm UV} < -20$ at different redshifts.
Unfortunately, there are no objects with  M$_{\rm UV} < -21$ at higher redshifts within the simulated volume and we cannot compare their virial masses at the brightest end.
Therefore, most of the evolution seen in \Fig{scalingR} is insensitive to the dust attenuation model.

For a fixed virial mass, a halo with $ {\rm log}(\Mv/\msun)=10.6$ at $z\simeq6$ typically hosts a relatively low luminosity galaxy with M$_{\rm UV} \simeq -19$. Another halo of the same mass at $z\simeq12$ contains a galaxy two magnitudes brighter, M$_{\rm UV} \simeq -21$.
This evolution at a given halo mass can also be due to a higher conversion of accreted gas into stars for a given halo growth at higher redshifts.
We will explore this possibility in the next section.



\subsection{Instantaneous galaxy formation efficiency}

   \begin{figure}
   \centering
   \includegraphics[width= \columnwidth]{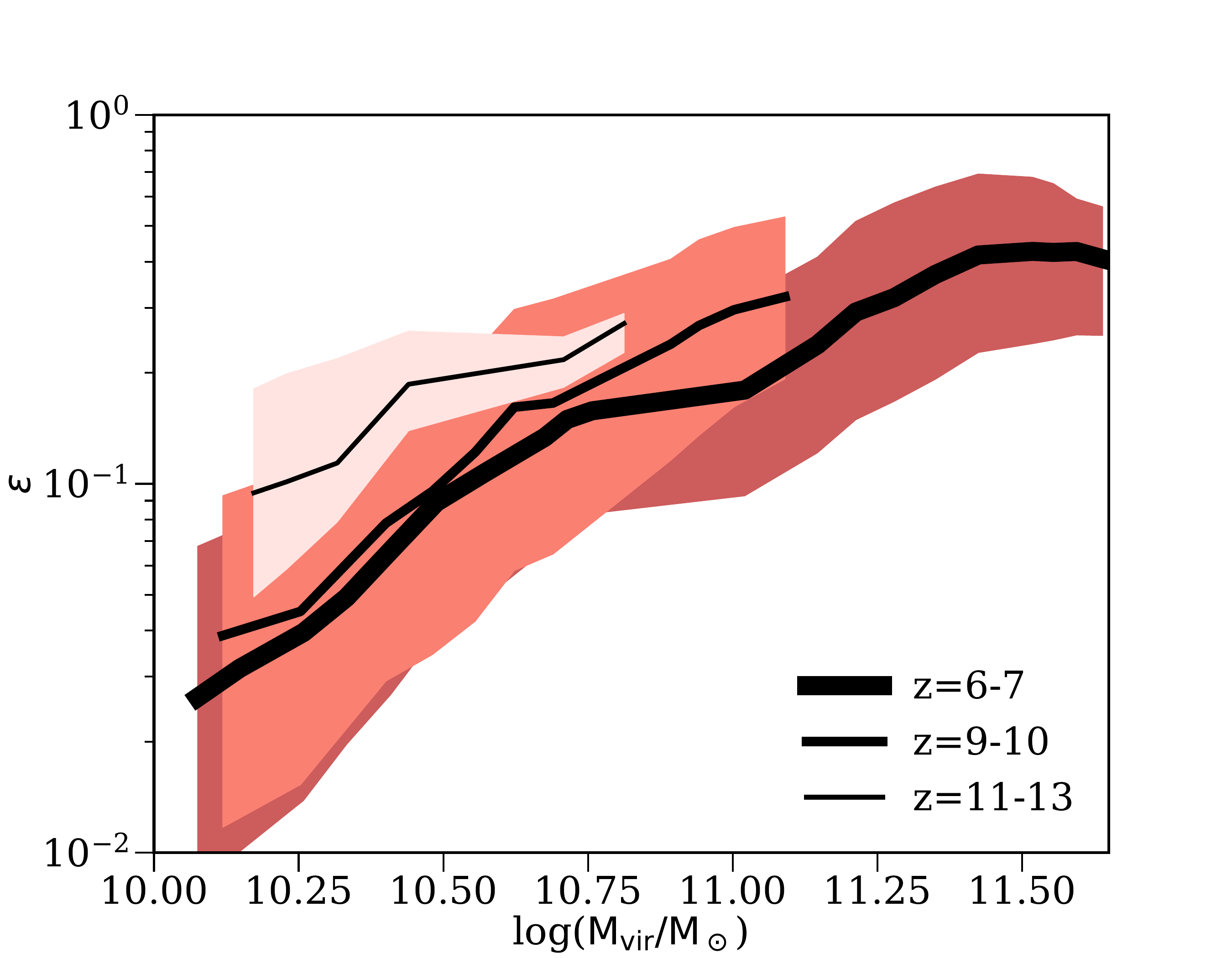}
   \caption{Instantaneous galaxy formation efficiency, $\epsilon=\dot{M}_{\rm *}/(\dot{M}_{vir} f_B)$, at different redshifts.
   The colored regions cover the 25 and 75\% quantiles. 
   The efficiency increases with mass and redshift.
   }
              \label{fig:epsilon}%
    \end{figure}

   \begin{figure}
   \centering
   \includegraphics[width= \columnwidth]{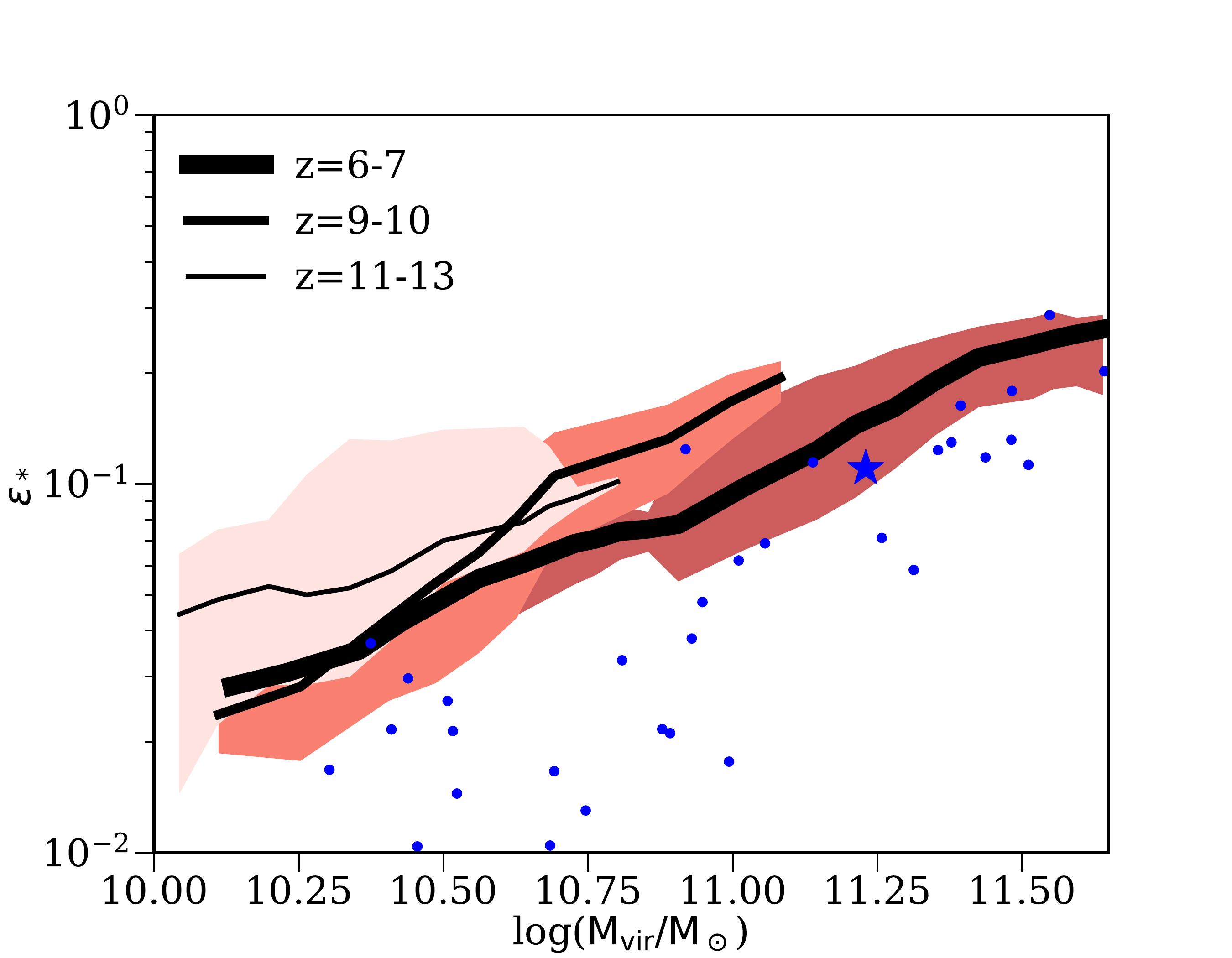}
   \caption{Integrated galaxy formation efficiency, $\epsilon_*=M_{\rm *}/(M_{vir} f_B)$, at different redshifts.
   The lines are the same as in \fig{epsilon}. Blue points represent the VELA-6 simulations \citep{Ceverino23} at $z=4$ 
   and the blue star represents a $z=0$ galaxy from \citet{Ceverino17}. 
   As redshift increases, the efficiency increases at a fixed virial mass. }
              \label{fig:epsilonS}%
    \end{figure}

   \begin{figure*}
   \centering
   \includegraphics[width=2 \columnwidth]{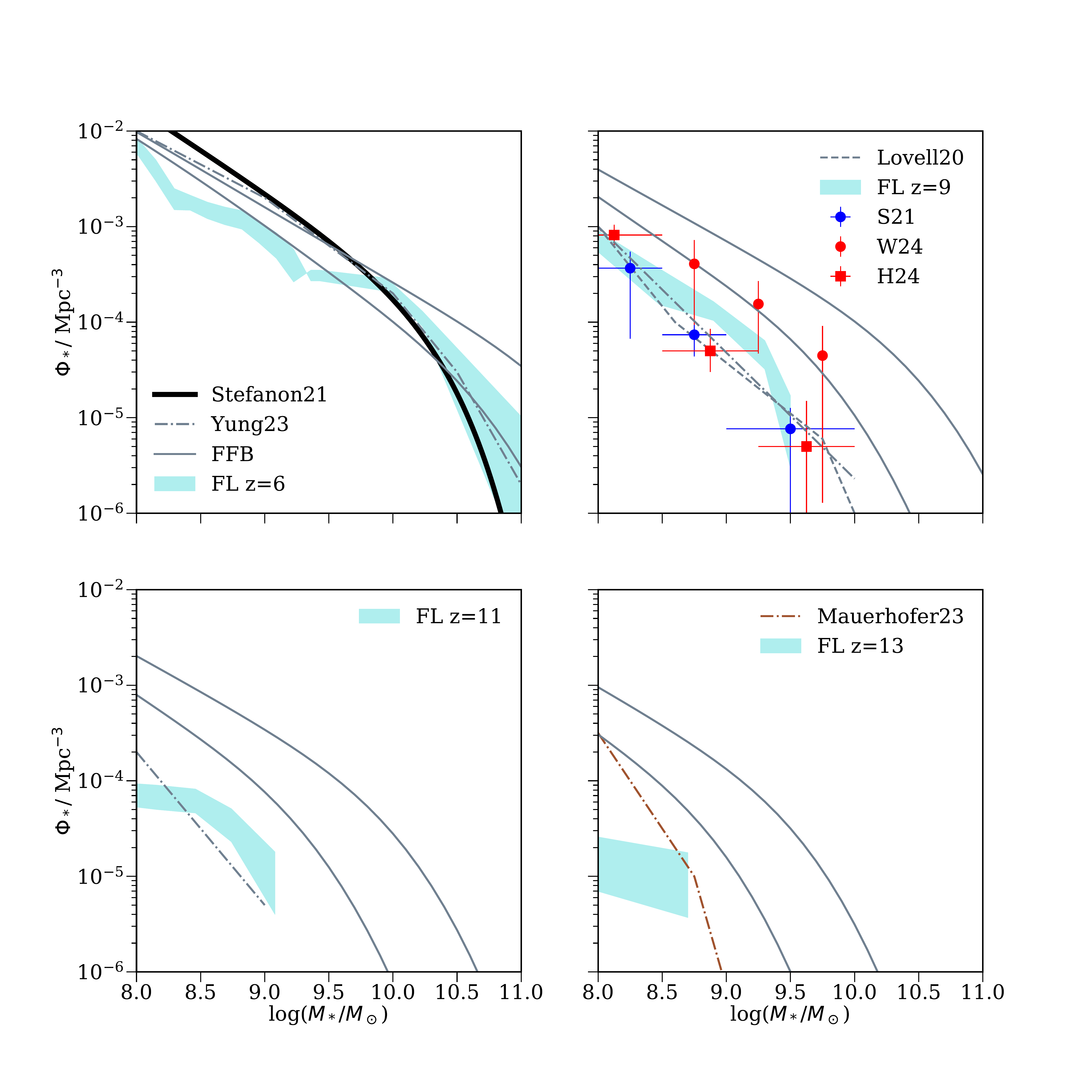}
   \caption{Evolution of the galaxy stellar mass  function from z=6 to z=13. 
   FirstLight is consistent with pre-JWST observations  \citep{Stefanon21} at $z=6$ but it shows higher numbers of massive galaxies at $z=9$, 
   in agreement with JWST results \citep{Harvey24, Weibel24}.
   Gray solid lines represents FFB model with $\epsilon_{\rm max}=0.2$ and 1.
   Simulations by \citet{Lovell21} and semi-analytical models by \citet{Yung23} and \citet{Mauerhofer23} show lower values at high masses.}
              \label{fig:GSMF}%
    \end{figure*}    

The instantaneous galaxy formation efficiency can be defined as the ratio between the stellar and halo mass growth,
\begin{equation} 
\epsilon= \frac{\dot{M}_{\rm *}}{\dot{M}_{\rm vir} f_B},
	\label{eq:epsilon}
\end{equation}
normalized to the universal baryonic fraction, $f_B$. 
This ratio removes the explicit redshift dependence, $(1+z)^{5/2}$, described above.
It mainly depends on the ability of accreted baryons to contribute to the galaxy stellar mass growth: the penetration of gas streams into the halo center and the regulation of the gas-stars cycle by feedback. These processes depend on halo mass. For example, feedback regulates strongly the formation of stars in low-mass halos with shallow potential wells \citep{DekelSilk86}. 
At a fixed halo mass, this efficiency may increase with redshift due to an increase in the penetration factor or due to a decrease in the mass-loading factor of galactic outflows \citep{Dekel14}. In the FFB scenario, this is due to an increase in the gas density \citep{Dekel23}. The galaxy efficiency can vary with redshift due to alterations in these processes.

We compute the instantaneous  efficiency using all snapshots stored every 10 Myr.  Mass variations, $\Delta \Ms$ and $\Delta \Mv$ between consecutive snapshots of the same galaxy can be compared within a given redshift bin. \Fig{epsilon} shows the evolution of this efficiency from $z\simeq6$ to 12. There is a significant increase of this efficiency with redshift.  At a fixed halo mass, $\Mv \simeq10^{11} \msun$, the efficiency increases a factor 3 between 10\% at $z\simeq6$ and $\sim$30\% at $z\simeq12$.
They are similar to the values required for reconciling UV and halo mass functions at $z\geq10$ \citep{Inayoshi22}. 
For more massive galaxies, $\Mv\simeq10^{11.5} \msun$, the values can be higher, although there is a flattening of the efficiency, with maximum values of $\sim$40\% at $z\simeq6$ on average. 
These massive galaxies are absent within the limited volume of these simulations at redshift higher than 7 and we cannot constrain the evolution at this massive end.    


\subsection{Integrated galaxy formation efficiency}

The integrated galaxy formation efficiency differs from the instantaneous efficiency because it involves time-integrated values:
\begin{equation} 
\epsilon_*= \frac{\Ms}{\Mv f_B},
\label{eq:epsilon*}
\end{equation}
where $\Ms$ is the stellar mass of the galaxy.
\Fig{epsilonS} shows the evolution of this efficiency with redshift. 
It has a similar dependence with virial mass than the instantaneous efficiency although its redshift evolution is milder.
At a fixed virial mass, $\Mv \simeq10^{11} \msun$, the integrated efficiency doubles its value from $\epsilon_*\simeq0.1$ at $z\simeq6$ to 
 $\epsilon_*\simeq0.2$ at $z\simeq10$. 
 At higher redshifts, $z\simeq12$, the apparent lack of evolution at this massive end is due to a small number of galaxies in this bin.
 At lower masses, the evolution is below a factor 2 and the efficiency remains at lower values,  $\epsilon_* < 0.1$, because feedback regulates efficiently the SF process at low masses.
 At a given redshift, the integrated efficiency is slightly lower than the instantaneous efficiency, \equ{epsilon}, by a factor of about 2. This is because the integrated efficiency is based on the time-integral of previous values of  $\dot{M}_{\rm *}/\dot{M}_{\rm vir}$, which are generally lower than the current value because of the galaxy growth and the mass-dependence described in \Fig{epsilon}.
 
 This redshift dependency continues toward lower redshifts.
Other simulations with the same feedback model show lower efficiencies at $z<5$.
 This is especially relevant at low masses, $\Mv < 10^{11} \msun$, where feedback is most efficient. 
For example, the VELA-6 simulations \citep{Ceverino23} have a factor 3-4 lower efficiency at $z=4$ than FirstLight galaxies with the same virial mass, $\Mv \simeq 10^{10.5} \msun$, at $z=12$.
On the other hand, the AGORA galaxy at $z\simeq0$ \citep{Ceverino17} has a more massive halo,  $\Mv \simeq 10^{11.2} \msun$ and it reaches the same efficiency, $\epsilon_*\simeq0.1$, than FirstLight galaxies of similar mass at $z\simeq6$. There are no counterparts at higher redshift and therefore we cannot see if the efficiency is higher at the high-mass end but the extrapolation from lower masses hints to that possibility.
We conclude that the integrated efficiency evolves with time, but this evolution depends on halo mass. Ultimately, it is related to the competition between the penetration of gas flows into galaxies and the outflows driven by feedback.
 

\subsection{Galaxy stellar mass function}

The evolution in the integrated galaxy formation efficiency drives a non-homogeneous evolution of the galaxy stellar mass function (\Fig{GSMF}).
This is consistent with recent JWST results at $z\simeq9$ \citep{Harvey24, Weibel24} within observational uncertainties.
Low-mass galaxies, $\Ms = 10^{8.5} \ \msun$, grow in numbers by a factor 5  between $z=11$ and $z=9$. Slightly more massive galaxies, $\Ms = 10^{9}  \ \msun$, grow by a factor 10 within the same period.
As a result, they are more abundant than in other models \citep{Lovell21,Yung23, Mauerhofer23} with lower efficiencies at  $z\ge10$.
FFB models with $\epsilon_{\rm max}=0.2$ show consistent results at $z=6$ but  higher number densities than FirstLight at $z\geq9$.

Galaxies more massive than $10^9 \ \msun$ at $z>11$ are absent within these cosmological volumes. 
Larger volumes are needed to simulate these massive and relatively rare halos. 
FirstLight also predicts higher number densities than pre-JWST observations \citep{Stefanon21} at $z\simeq9$, even after different assumptions on IMF and cosmology are corrected. 
However, this excess almost disappears at $z\simeq6$, as the integrated efficiency decreases with time.

The lower numbers found at low masses indicate some incompleteness in the small and intermediate boxes at  the lowest redshifts, $z\simeq6$ \citep{PaperII}.  
However, the agreement between observations and FirstLight at high masses, $\Ms > 10^{10} \msun$, is  good within uncertainties.
There are ways to correct for incompleteness \citep{Ma18, Lovell21} but they are not used here, as we are focusing on high masses and redshifts, where FirstLight includes all halos available  within  the simulated volumes.


\section{Discussion}
\label{sec:discussion}

\subsection{Comparison of integrated efficiency from observations and other models}

   \begin{figure}
   \centering
   \includegraphics[width= \columnwidth]{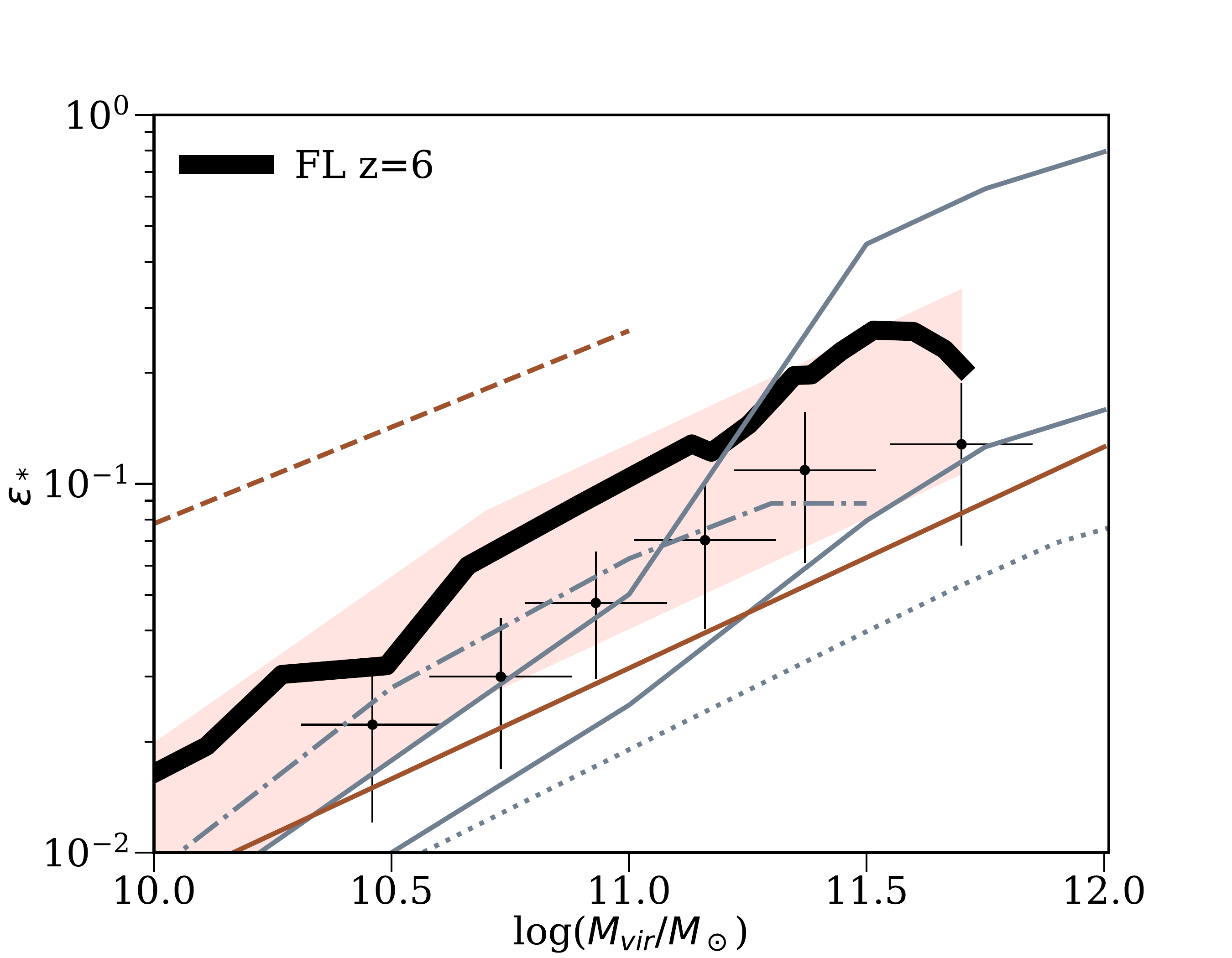}
      \includegraphics[width= \columnwidth]{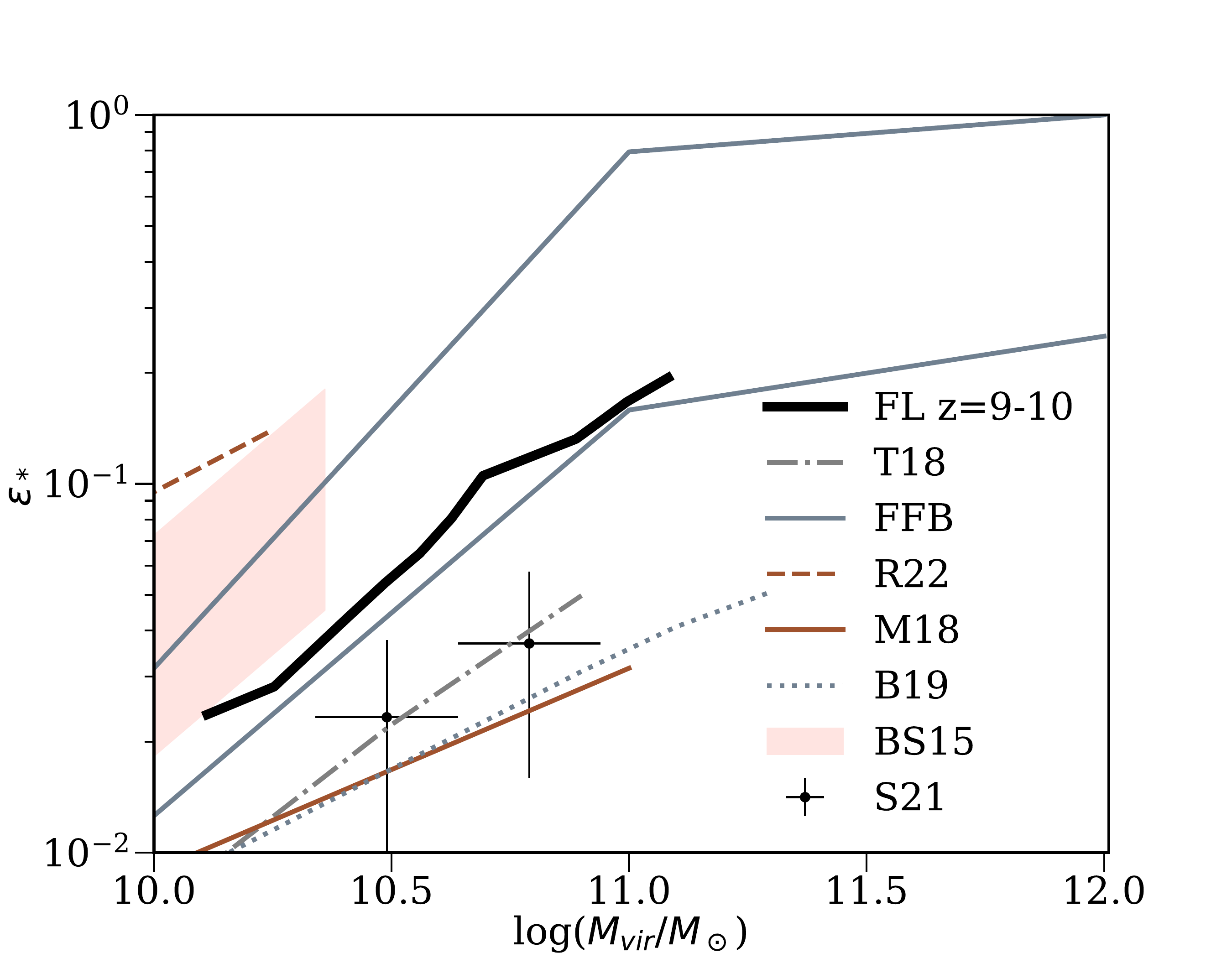}
   \caption{Comparison of the integrated efficiency at $z=6$ (top) and $z\simeq9$ (bottom) between FirstLight (solid black lines, as in \Fig{epsilonS}),  pre-JWST observations by \cite{Stefanon21} (points with error bars), SPHINX (dashed) and FIRE-2 (solid) simulations, and semi-empirical models by \cite{BehrooziSilk15} (red contour), \cite{Tacchella18} (dot-dashed), \cite{Behroozi19} (dotted), and \cite{LiDekel23}  models of FFB with $\epsilon_{\rm max}=0.2$ and 1 (solid grey lines).}
              \label{fig:comp}%
    \end{figure}

The integrated efficiency, \equ{epsilon*}, has been computed in different works but there is no consensus on whether it is time dependent or independent. 
\Fig{comp} shows a comparison between FirstLight results and other works at two different redshifts.
\cite{Stefanon21} use the deepest Spitzer and HST observations and show that the integrated efficiency increases with mass but there is no significant evolution from $z\simeq10$ to $z\simeq6$. 
FirstLight is consistent with their results at $z=6$ but it is significantly higher at $z\simeq9$ by a factor 2-4.
Their  low values are mostly driven by the accelerated evolution of the galaxy stellar mass function between $z\simeq9$ and $z\simeq6$ (\Fig{GSMF}), inconsistent with the FirstLight results.
The semi-empirical model by \citet{Tacchella18}  combine UVLF from HST observations at $z=5-10$ and dark-matter halo accretion rates from N-body simulations. They also conclude that the integrated efficiency is redshift-independent in that redshift range.
Therefore, they agree with FirstLight at $z=6$ but they are significantly lower at $z=9$, in agreement with the previous observations.
A similar approach is used in others abundance-matching models \citep[e.g.][]{Behroozi19} that use pre-JWST observations to infer relatively low efficiencies at high redshifts.

Other cosmological simulations, like SPHINX \citep{Rosdahl18, Katz23}, have a high efficiency. It therefore over-predicts the values at low redshifts, $z=6$, where observational estimates are more robust. 
On the other hand, the FIRE-2 simulations \citep{Ma18} under-predict the efficiency at the same redshift. 
Both simulations have roughly constant values with time because feedback is either too strong (FIRE) or too weak (SPHINX) to self-regulate the combined cycle of SF and feedback.

Other models consider a redshift-dependent efficiency, which is qualitatively consistent with FirstLight.
For example, \cite{BehrooziSilk15} use a variable ratio between the sSFR and the halo specific mass accretion rate to predict the galaxy efficiency to $z\simeq15$. 
Although their estimates are higher than in previous redshift-independent models, they are consistent with FirstLight at $z=6$.
At  $z\simeq10$,  their uncertainties are higher and these predictions are also above FirstLight by a factor of a few. 

In the FFB scenario \citep{Dekel23}, the efficiency is a function of mass and redshift, assuming a maximum value, $\epsilon_{\rm max}$, above a mass threshold, ${\rm M}_{\rm vir,fbb}$, that declines with redshift \citep{LiDekel23}. 
For halo masses below that threshold, feedback  regulates the SF cycle and maintains a relatively low SF efficiency.
Their minimum model with $\epsilon_{\rm max}=0.2$ is consistent with FirstLight at $z=10$.
Order-unity efficiencies, $\epsilon_{\rm max}\simeq1$, are excluded within the mass range explored by FirstLight.
At $z=6$, the maximum efficiency in FirstLight is close to $\epsilon_*\simeq0.3$.
This is consistent with the cumulative values estimated using MIRI photometry of massive galaxies \citep{Wang24}.
This maximum happens at $\Mv\simeq10^{11.5} \ \msun$, slightly below the FFB threshold,  ${\rm M}_{\rm vir,fbb}\simeq10^{11.75} \ \msun$ at $z=6$.
As a result, the FirstLight efficiencies at $z=6$ are higher than in the FFB scenario at lower masses,  $\Mv < 10^{11} \ \msun$, a mass range in which the efficiencies in \citet{LiDekel23} get closer to the \citet{Behroozi19} model, which shows low values.


 
\subsection{Evolution of the cosmic UV and stellar mass density}   

   \begin{figure}
   \centering
   \includegraphics[width= \columnwidth]{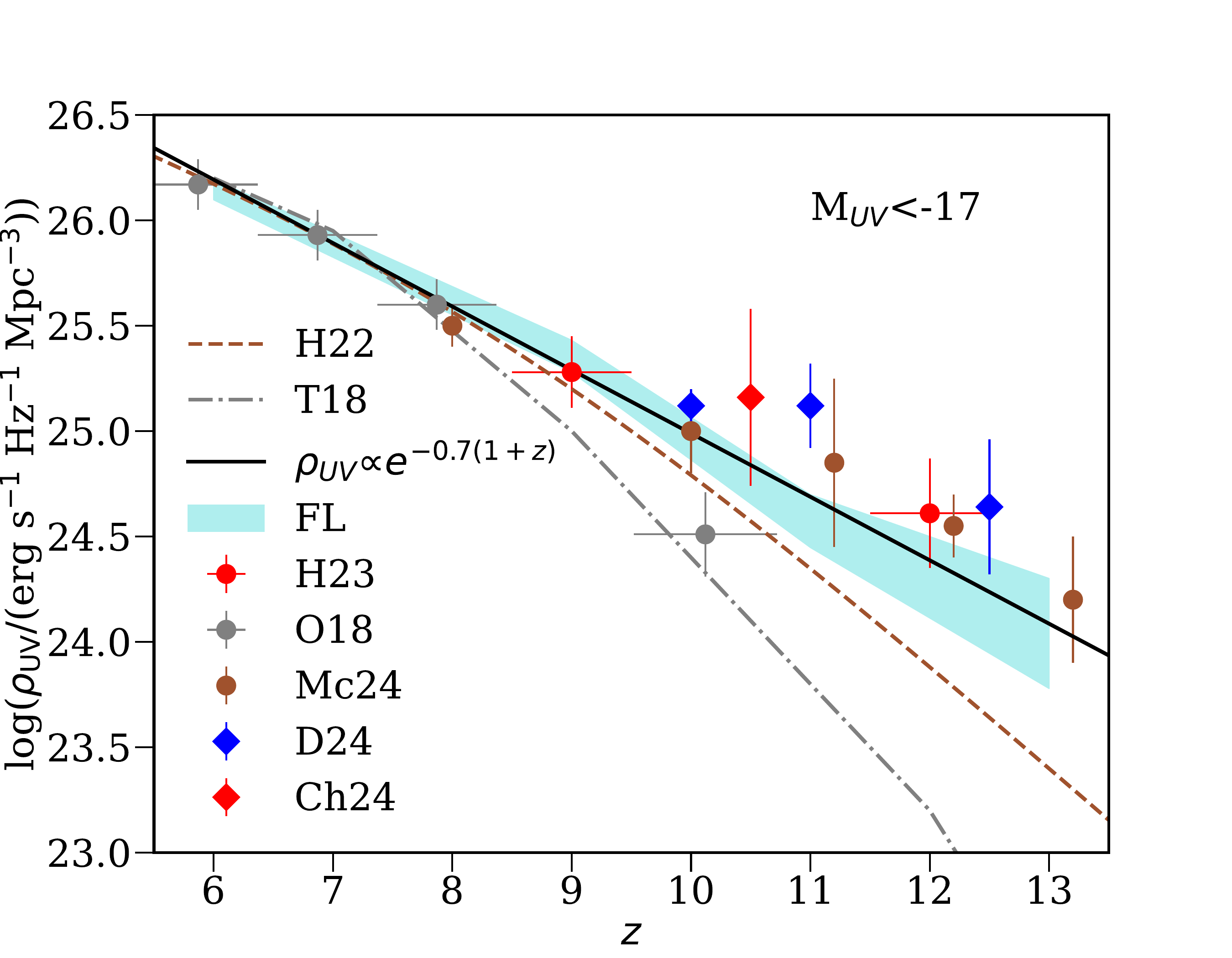}
   \caption{Evolution of the cosmic UV density for galaxies with $M_{UV}<-17$.
   FirstLight values are above the models with redshift-independent efficiency by \citet{Harikane22} and \citet{Tacchella18}, as well as other pre-JWST observations at $z\geq10$ \citep{Oesch18}.
   JWST values \citep{Harikane23, McLeod24,Chemerynska24, Donnan24} are consistent with FirstLight within the estimated uncertainties.
   }
              \label{fig:cosmicUV}%
    \end{figure} 
  
  The integration of the UVLF (\Fig{UVLF}) to   M$_{\rm UV}<-17$ at different redshifts gives an estimate of the comoving cosmic UV density (\Fig{cosmicUV}), which can we compared with other works.
  Due to the increase of the efficiency at $z\ge9$, the values are above the models with redshift-independent efficiency \citep{Tacchella18, Harikane22}.
  There is no sign of accelerated evolution between $z=10$ and $z=8$, as seen in pre-JWST observations \citep{Oesch18}.
  In fact, a gentle evolution, approximately fitted by log($\rho_{\rm UV}) \propto -0.7 (1+z)$, is consistent with most  JWST observations up to $z\simeq13$.
  FirstLight results  are only slightly below observations by 0.2-0.3 dex at $z\geq11$, still in between the estimated uncertainties. Larger samples of galaxies with robust redshift estimations are necessary for more accurate constrains.   
  
    
   %
   \begin{figure}
   \centering
   \includegraphics[width= \columnwidth]{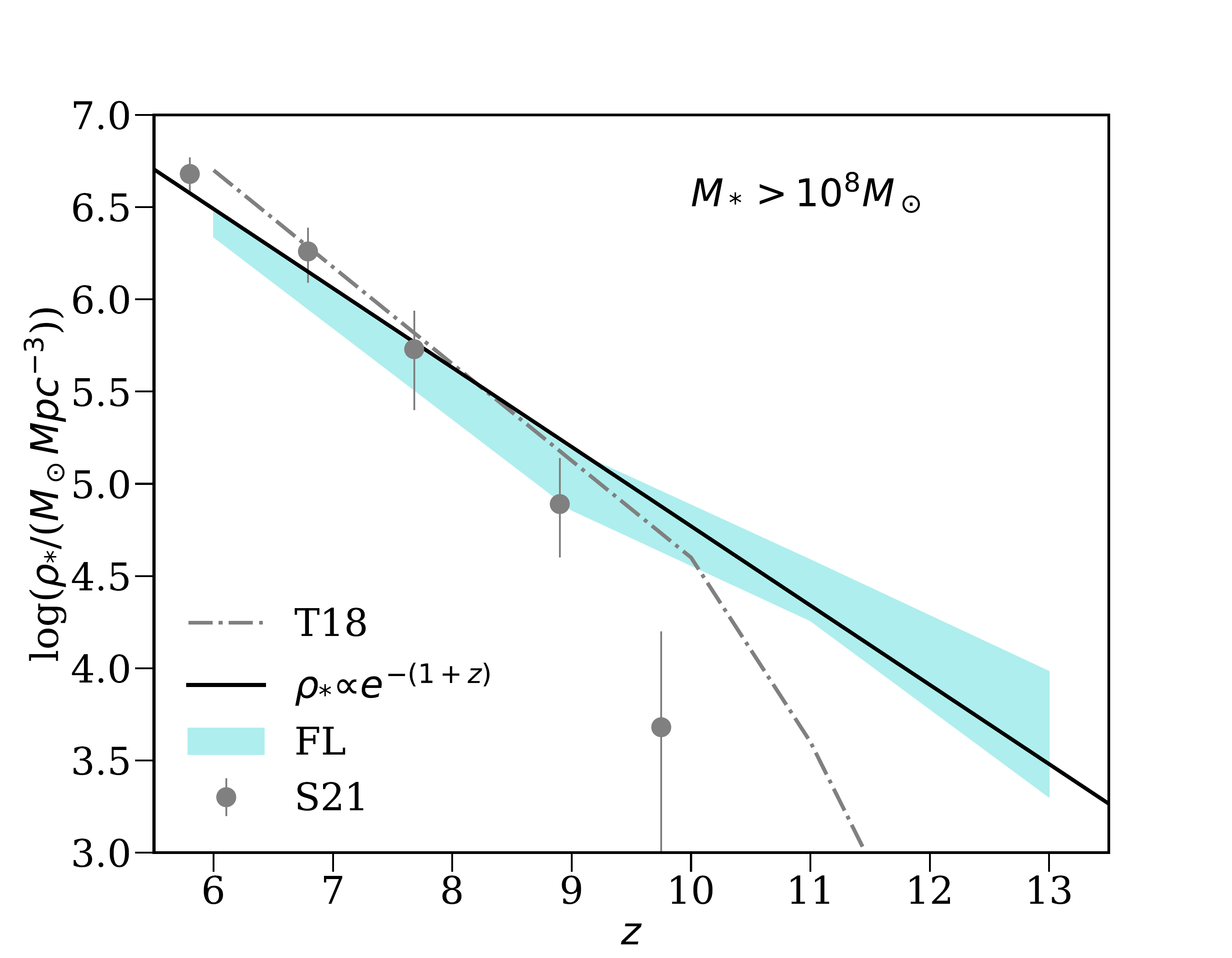}
   \caption{Evolution of the cosmic stellar mass density for galaxies with $\Ms>10^8 \msun$. 
   FirstLight-derived densities are higher than pre-JWST values at $z>9$ \citep{Stefanon21} and previous models with a redshift-independent efficiency \citep{Tacchella18}. }
              \label{fig:cosmicMs}%
    \end{figure}   
    
The comoving cosmic stellar mass density, for all galaxies more massive than  $10^8 \msun$, provides another overall comparison with previous works (\Fig{cosmicMs}).
At $z\simeq6$, FirstLight values are slightly lower than other models. This is due to the slight incompleteness at low masses discussed above.
At  higher redshifts, $z>9$, models that assume a constant efficiency \citep{Tacchella18} and pre-JWST observations \citep{Stefanon21} show a much more drastic drop due to the lack of massive galaxies.
However, FirstLight values show a smooth evolution,   log($\rho_{*}) \propto -(1+z)$, due to the high efficiency at these high redshifts.
Future JWST observations are needed to confirm the gentle rise of the stellar mass density at these early times.


\subsection{Drivers of galaxy efficiency}    

   \begin{figure}
   \centering
   \includegraphics[width= \columnwidth]{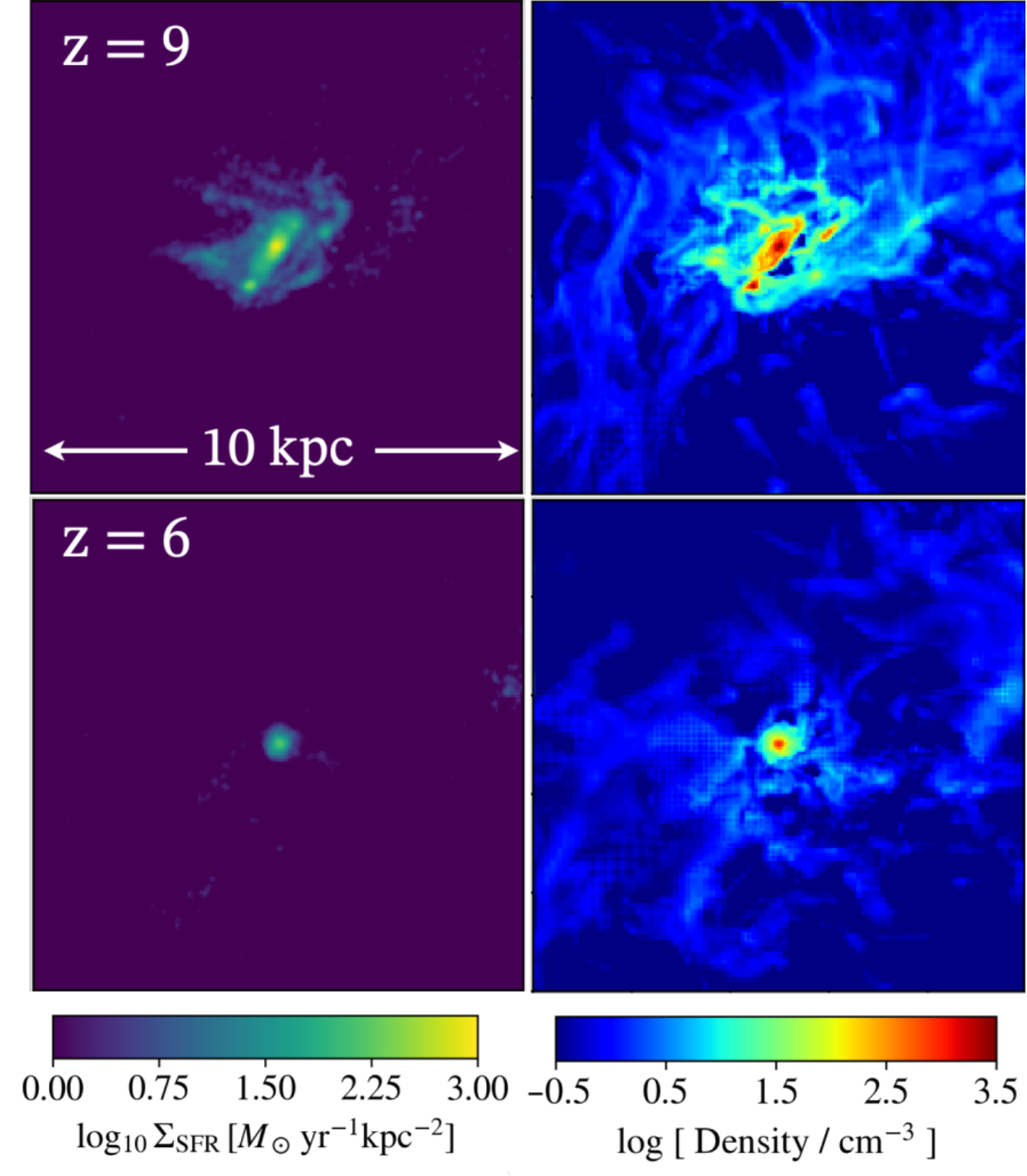}
   \caption{Example of a highly efficient galaxy at $z=9$ (top) and a galaxy at $z=6$ with the same halo mass of $\Mv\simeq 10^{11} \msun$ but lower efficiency (bottom). Left panels show the star formation rate surface density and right  panels depict the gas density, averaged along the line of sight.}
              \label{fig:example}%
    \end{figure}

Different galaxy conditions determine different efficiencies in the galaxy growth with respect to halo growth, \equ{epsilon}.
For example, the baryonic inflow from the halo to the galaxy depends on the cycle from the circumgalactic medium to the star-forming interstellar medium.
Variations in this cycle could modify the rate of galaxy growth even at a fixed halo mass growth \citep{Dekel13}. 
Feedback regulates the SF process by driving galactic outflows or by decreasing the fraction of cold and dense, star-forming gas \citep{Ceverino14}. 
However, feedback depends on gas conditions, particularly on its density \citep{DekelSilk86, Dekel23}.
Therefore, galaxy efficiency can vary with redshift as gas within galaxies evolves with time.
A more in-depth study to identify the main driver of this evolution is beyond the scope of this paper, but we can provide some hints.

In \Fig{example} we compare two examples of galaxies with a similar halo mass, $\Mv\simeq 10^{11} \msun$, but different galaxy efficiencies at different redshifts. The first galaxy at $z=9$ has a high efficiency, $\epsilon\simeq0.3$ and $\epsilon_*\simeq0.6$. 
A significant fraction of the galaxy, especially the center and off-center clumps, has very high densities, $n \ge 3000 \ {\rm cm}^{-3}$. The corresponding free-fall time is lower than 1 Myr and  the effect of feedback drastically decreases at these high densities, according to the FFB scenario \citep{Dekel23}.
Therefore, the SFR surface density reaches extreme values, $\Sigma_{\rm SFR} \ge 10^3 \msun {\rm yr}^{-1} \kpc^{-2}$, 
similar to the observational estimates in a very dense star-forming environment of a $z=12$ galaxy \citep{Calabro24}.

The second example at $z=6$ has a lower efficiency, $\epsilon\simeq0.1$ and $\epsilon_*\simeq0.09$. Both the gas density and  the SFR surface density are much lower than in the previous example, 
although consistent with observations that spatially resolve the Kennicutt-Schmidt relation at $z=7$ \citep{Vallini24}.
This galaxy has the same amount of gas within the virial radius as in the previous case ($M_{\rm gas, halo}=2 \times 10^{10} \ \msun$), but only 7\% of this mass is forming stars. This fraction increases to 30\% in the first galaxy at $z=9$. 
As a result, the star-formation rate of this highly efficient galaxy is ten times higher than the other example at lower redshift. 
This also translates into a factor-of-two shorter gas depletion time, $t_{\rm D}=M_{\rm gas, galaxy}/{\rm SFR}=60  \Myr$. 
This value is close to the average at $z=10$ \citep{PaperII}, but it  is a factor 30 shorter than in local galaxies \citep{Leroy08, Saintonge22, SunLeroy23}.
We conclude that the gas densities in SF bursts at extremely high redshifts, $z\ge9$, facilitates very high galaxy efficiencies, $\epsilon_{\rm max}\simeq 0.3-0.4$, at moderate halo masses, $\Mv \simeq 10^{11} \ \msun$. Future simulations of bigger volumes will explore higher masses and possible higher efficiencies in more extreme and rare halos at high redshifts.

    
\section{Conclusions}
\label{sec:conclusions}

We use the FirstLight database of 377 zoom-in cosmological simulations of a mass-complete sample of galaxies with a spatial resolution of 10-20 parsecs in a $(60 \ {\rm Mpc})^3$ comoving volume,
 to understand galaxy growth at extremely high redshifts, $z=9-13$. 

The main highlights of this paper are the following: 
   \begin{enumerate}
      \item FirstLight contains a high number of bright galaxies, M$_{\rm UV} \simeq -20$, consistent with JWST data.
      \item The relation between  M$_{\rm UV}$ and halo mass evolves with redshift. It is mostly independent  of the dust attenuation model and it is due to higher mass accretion and SF efficiency at higher redshifts.
      \item The galaxy formation efficiency, \equs{epsilon} and \equm{epsilon*}, increases significantly with mass and redshift. At a fixed mass, galactic halos at extremely high redshifts convert gas into stars at a higher rate than at lower redshifts.
      \item The number density of massive galaxies, $\Ms \simeq 10^{9} \ \msun$ at $z\ge9$, are higher than in other models with a constant star-formation efficiency.
      \item The high gas densities in these SF bursts at  $z\ge9$ enable these high efficiencies (\Fig{example}), in qualitative agreement with the scenario of Feedback Free Starbursts (FFB) at high redshifts.
   \end{enumerate}

In spite of this success in reproducing current JWST observations, FirstLight simulations also have some caveats. 
For example, they do not include the radiative transfer of ionizing photons, instead assuming that these photons are mostly absorbed in the immediate
      vicinity of the massive stars producing them. Therefore, radiative feedback only affects gas close to massive stars; the effects of the leakage of this
      radiation into the interstellar medium is not included.
 This could modify the gas conditions further away from the star-forming regions \citep{Emerick19}. Future simulations should include these radiative-transfer effects.
 
 The maximum resolution achieved in FirstLight (10-20 pc) also imposes some limitations. 
 This is not enough to properly resolve the star forming clouds that are predicted to have comparable sizes of $\sim$10 pc. The feedback-free starbursts with order-unity SF efficiency are predicted to occur in these clouds because of their high gas densities. 
 High-resolution simulations of the formation of individual massive clusters find these high SF efficiencies in high-density clouds. 
 \citep[e.g.][]{Kim16, Fukushima21, Calura22, Polak23}.
  In the FirstLight simulations, similar high densities happen on larger scales at higher redshifts (\Fig{example}) because of the compact nature of these galaxies. Therefore, the efficiency increases with redshift but its maximum value, $\epsilon_{\rm max}\simeq 0.2-0.3$, is far away from the order-unity efficiency expected within feedback-free starbursts. Still, this galaxy-averaged efficiency is consistent with the fiducial FFB model that best agree with JWST observations \citep{LiDekel23}.  
  
 Other cosmological simulations (Rosdahl et al. 2018; Ma et al. 2018) do not show this evolution of the efficiency because their feedback is either too weak or too strong to self-regulate the combined cycle of SF and feedback at high-z. The model included in FirstLight has the combined effect of thermal, kinetic and radiative feedback. At high densities, both thermal and kinetic modes are quickly dissipated by shocks and radiative cooling. This reduces the effect of feedback in high densities conditions at high redshifts.

We conclude that FirstLight  simulations can reproduce different conditions of galaxy formation across cosmic history. Relatively massive galaxies at extremely high redshifts have a low efficiency of feedback and therefore star formation and galaxy growth can proceed faster than at later times. Other modern cosmological simulations with phenomenological feedback models \citep{Pillepich18, Dave19} include a decrease of the mass in galactic outflows with increasing galaxy mass. However, these models are calibrated on observations at lower redshifts and they may fail at much early times, when gas densities are much higher.
Future observations by JWST will utilize multi-line diagnostics
to infer the gas density in star-forming regions at these high-z and they can validate the redshift evolution advocated here.

The next generation of cosmological simulations must reproduce the galaxy conditions across different redshifts, from the local Universe at $z=0$ well into the dark epoch, $z=10-20$, before the reionization of the Universe.
Parsec scale resolution is probably required in order to properly resolve star formation and self-regulation of stellar birth at this epoch, including key processes such as the turbulence in the dense interstellar medium, feedback from black holes and AGN, as well as magnetic fields and cosmic rays.


\begin{acknowledgements}
     We thank the referee, Avishai Dekel, for constructive suggestions that improve the quality of this paper. We acknowledge stimulating discussions with Joel Primack and Anatoly Klypin.
     We thank Viola Gelli and Zhaozhou Li for sharing their data.
     The authors gratefully acknowledge the Gauss Center for Supercomputing for funding this project by providing computing time on the GCS Supercomputer SuperMUC at Leibniz Supercomputing Center (Project ID: pr92za). 
     The authors thankfully acknowledge the computer resources at MareNostrum and the technical support provided by the Barcelona Supercomputing Center (RES-AECT-2020-3-0019).
     This work used the v2.1 of the Binary Population and Spectral Synthesis (BPASS) models as last described in Eldridge et al. (2017).
     DC is a Ramon-Cajal Researcher and is supported by the Ministerio de Ciencia, Innovaci\'{o}n y Universidades (MICIU/FEDER) under research grant PID2021-122603NB-C21.
     YN acknowledges funding from JSPS KAKENHI Grant Number 23KJ0728 and a JSR fellowship.      
     RSK and SCOG acknowledge funding from the ERC via Synergy Grant "ECOGAL" (project ID 855130), from the German Excellence Strategy via the Heidelberg Cluster of Excellence (EXC 2181 - 390900948) "STRUCTURES", and from the German Ministry for Economic Affairs and Climate Action in project ``MAINN'' (funding ID 50OO2206). RSK and SCOG also thank for computing resources provided by the Ministry of Science, Research and the Arts (MWK) of the State of Baden-W\"{u}rttemberg through bwHPC and DFG through grant INST 35/1134-1 FUGG and for data storage at SDS@hd through grant INST 35/1314-1 FUGG.
\end{acknowledgements}

%
\bibliographystyle{aa} 
 \bibliography{FLV_v5} 
%

\end{document}